\def\a{\alpha}\def\b{\beta}\def\c{\chi}\def\d{\delta}\def\e{\epsilon}
\def\f{\phi}\def\h{\theta}
\def\l{\lambda}\def\m{\mu}\def\n{\nu}\def\q{\psi}\def\r{\rho}
\def\y{\eta}

\def\L{\Lambda}

\def\de{\partial}
\def\inf{\infty}\def\mo{{-1}}\def\ha{{1\over 2}}

\def\({\left(}\def\){\right)}\def\[{\left[}\def\]{\right]}
\def\lra{\leftrightarrow}

\def\const{{\rm const}}

\def\mn{{\mu\nu}}

\def\tran{transformations }\def\coo{coordinates }

\def\rep{representation }

\def\cor{commutation relations }

\def\section#1{\bigskip\noindent{\bf#1}\smallskip}
\def\subsect#1{\bigskip\noindent{\it#1}\smallskip}

\def\PL#1{Phys.\ Lett.\ {\bf#1}}

\def\PR#1{Phys.\ Rev.\ {\bf#1}}\def\CQG#1{Class.\ Quantum Grav.\ {\bf#1}}
\def\NP#1{Nucl.\ Phys.\ {\bf#1}}
\def\JMP#1{J.\ Math.\ Phys.\ {\bf#1}}

 \def\IJMP#1{Int.\ J. Mod.\ Phys.\ {\bf #1}}

\def\JHEP#1{JHEP\ {\bf#1}}
\def\RMP#1{Rev.\ Mod.\ Phys.\ {\bf#1}}\def\AdP#1{Annalen Phys.\ {\bf#1}}
\def\arx#1{{\tt arXiv:#1}}

\def\ref#1{\medskip\everypar={\hangindent 2\parindent}#1}
\def\beginref{\begingroup
\bigskip
\centerline{\bf References}
\nobreak\noindent}
\def\endref{\par\endgroup}

\def\hP{\hat P}\def\hX{\hat X}\def\hL{\hat L}\def\hJ{\hat J}
\def\hp{\hat p}\def\hx{\hat x}\def\hA{\hat A}\def\hK{\hat K}
\def\state{|j,l,m\rangle}
\def\vf{\varphi}\def\ij{{ij}}


\magnification=1200

{\nopagenumbers
\line{}
\vskip60pt
\centerline{\bf Quantum mechanics on a curved Snyder space}
\vskip60pt
\centerline{
{\bf S. Mignemi}\footnote{$^\ddagger$}{e-mail: smignemi@unica.it}
and {\bf R. \v Strajn}\footnote{$^\dagger$}{e-mail: ri.strajn1@studenti.unica.it}}
\vskip10pt
\centerline {Dipartimento di Matematica e Informatica, Universit\`a di Cagliari}
\centerline{viale Merello 92, 09123 Cagliari, Italy}
\smallskip
\centerline{and INFN, Sezione di Cagliari}
\vskip80pt
\centerline{\bf Abstract}
\medskip
{\noindent We study the representations of the three-dimensional Euclidean Snyder-de Sitter
algebra. This algebra generates the symmetries of a model admitting two fundamental scales
(Planck mass and cosmological constant) and is invariant under the Born reciprocity for
exchange of positions and momenta.
Its representations can be obtained starting from those of the Snyder algebra, and exploiting
the geometrical properties of the phase space, that can be identified with a Grassmannian
manifold.
Both the position and momentum operators turn out to have a discrete spectrum.}
\vskip10pt
{\noindent

}
\vskip80pt\
\vfil\eject}

\section{1. Introduction}

The Snyder-de Sitter (SdS) model, or triply special relativity [1], was introduced as
a generalization of the Snyder model [2] to a curved background.

The Snyder model has been the first example of noncommutative geometry proposed in the
literature, and is based on a deformation of the Heisenberg algebra by a fundamental
invariant scale $\b$ with dimension of inverse energy. It was introduced in the
hope of achieving a regularization of field theory through the new scale, whose
presence, contrary to what one may expect, does not affect the Lorentz invariance,
but only deforms the translation symmetry [3]. As in other cases of deformed Poincar\'e
invariance [4], the momentum space can be identified with a nontrivial manifold, namely
a 3-sphere $S^3$.

The properties of the Snyder space and its dynamics, both in the nonrelativistic and
relativistic version, have been investigated in several papers and in various contexts,
both classical and quantum [5,6,7,8,9].
In particular, it was shown that space is discretized [6], and deformed Heisenberg uncertainty
relations hold, implying a lower bound on measurable length [5].
Both classical and quantum dynamics are modified with respect to the standard results, with
deviations of order $\b^2E^2$, $E$ being the energy of the system [5].

The extension of the Snyder model to a spacetime background of constant curvature was
proposed in [1] (see also [10] for a different approach). This generalization was motivated
by the necessity
of including the cosmological constant $\L\sim\a^2$ among the bare parameters of a theory of
quantum gravity [1].
In this way, one introduces a third fundamental constant besides the speed of light $c$ and
the Snyder parameter $\b$, whence the name of triply special relativity originally given to
the theory.
The most relevant feature of this generalization is its duality for the interchange between
positions and momenta, that realizes Born reciprocity principle [11].
The properties of SdS space were studied by many authors [12,13,14], mainly in its
nonrelativistic version. In this case, both positions and momenta have discrete spectra, and
a minimal momentum occurs besides minimal length.

In this paper, we attempt to generalize the results on the algebraic structure of the
three-dimensional nonrelativistic Snyder model, investigated in [6], to the case of a
curved background. Although the nonrelativistic limit is physically less interesting
than the relativistic theory, it can shed some light on the structure of the theory, and
serve as a first step in the construction of a quantum field theory. Of course, in this limit
only two fundamental constants, $\a$ and $\b$, are left.

In the SdS case, the algebraic structure is less useful than for the Snyder case, because
for SdS the spectrum of the position
operator cannot be derived directly from the algebraic structure of the theory.
Using the relation between the representations of Snyder and of SdS found in [13],
it is however possible to find analytically the spectrum of the square of the position and
momentum operators.
Because of the Born reciprocity these spectra are essentially identical.

We also show that the SdS phase space can be identified with a Grassmannian coset space
$Gr(3,5)=SO(5)/SO(3)\times SO(2)$. This property may be exploited to construct deformed
addition laws for momenta, following the approach of [8].

\subsect{1.1 The SdS model}

The nonrelativistic SdS algebra $\cal A$  depends on two parameters $\a$ and $\b$ and contains
the usual
generators of rotations $\hL_k=\ha\e_{ijk}\hL_{ij}$ ($i,j=1,\dots,3$), with their standard
action on the position and momentum operators $\hx_i$ and $\hp_i$:
$$[\hL_i,\hL_j]=i\e_{ijk}\hL_k,\qquad
[\hL_i,\hx_j]=-i\e_{ijk}\hx_k,\qquad
[\hL_i,\hp_j]=-i\e_{ijk}\hp_k,\eqno(1.1)$$
The \cor of $\hx_i$ and $\hp_i$ satisfy instead a deformation of the Heisenberg algebra,
$$[\hx_i,\hx_j]=i\b^2\e_{ijk}\hL_k,\qquad[\hp_i,\hp_j]=i\a^2\e_{ijk}\hL_k,,$$
$$[\hx_i,\hp_j]=i\big[\d_{ij}+\a^2\hx_i\hx_j+\b^2\hp_j\hp_i+\a\b(\hx_j\hp_i+\hp_i\hx_j)\big],
\eqno(1.2)$$
where $\hL_{ij}=\ha(\hx_i\hp_j+\hp_j\hx_i-\hx_j\hp_j-\hp_i\hx_j)$.

For special values of the parameters, $\cal A$ reduces to the nonrelativistic Snyder algebra
($\a=0$), or the algebra of isometries of $S^3$ in Beltrami \coo ($\b=0$). It is possible to
define also a noncompact version of the algebra,
by analytic continuation to imaginary values of $\a$ and $\b$, with rather different properties
[13], but we shall not discuss it here.
As mentioned above, the algebra is invariant for $\a x_\m\lra\b p_\m$. More generally,
it is invariant for rotations in the phase space, $\a x_i\to\a x_i\cos\h+\b p_i\sin\h$,
$\b p_i\to-\a x_i\sin\h+\b p_i\cos\h$. Of course, this symmetry holds also in standard
quantum mechanics, with $\a=\b=1$.

The SdS algebra can be considered as a nonlinear realization of a model proposed by
Yang [15], which differs from SdS only in the assumption of a standard Heisenberg algebra
for positions and momenta, $[\hx_i,\hp_j]=i\hK\d_{ij}$, $\hK$ being a central charge
for the rotation group, satisfying $[\hK,\hx_i]=i\a^2\hp_i$, $[\hK,\hp_i]=-i\b^2\hx_i$.
With the identifications $\hL_{ij}=\hJ_\ij$, $\a\hx_i=\hJ_{4i}$, $\b\hp_i=\hJ_{5i}$,
$\hK=\hJ_{45}$, the Yang model reproduces an $so(5)$ algebra with generators $\hJ_\mn$,
($\m,\n=1,\dots,5$).

Also the 3-dimensional nonrelativistic SdS model enjoys an $SO(5)$ symmetry. In fact, its
phase space can be realized on the six-dimensional Grassmannian coset space
$Gr(3,5)=SO(5)/SO(3)\times SO(2)$,
with $SO(3)$ generated by the $J_{ij}$ and $SO(2)$ by $J_{45}$.

The space $Gr(3,5)$ can be parametrized by homogeneous \coo $x_\m$ and $p_\m$,
that satisfy the constraints [16]
$$\a^2x_\m^2=1,\qquad\b^2p_\m^2=1,\qquad x_\m p_\m=0.\eqno(1.3)$$
This parametrization associates a one-parameter set of matrices to each coset.
One can then identify the variables $x_\m$ and $p_\m$ with canonical \coo of a ten-dimensional
phase space and hence reduce it to a six-dimensional phase space parametrized by $x_i$ and
$p_i$ by eliminating the constraints (1.3), using the Dirac formalism [17].
In order to obtain a one-to-one parametrization, one has however to impose a further
constraint on the parameters $x_4$, $x_5$, $p_4$, $p_ 5$. This is also required by the Dirac
formalism since the constraints (1.3) split into one first class and two second class
constraints [12]. Unfortunately, not every constraint leads to the SdS algebra, and
one has therefore to choose a suitable gauge [12]. In particular, the choice
$\a x_5+\b p_5=0$ yields the algebra (1.2).

\section{2. The nonrelativistic Snyder algebra}

In this section we review some results on the \rep of the Snyder algebra from [5] and [6],
that will be useful in the following discussion.

The Snyder algebra is the limit of the SdS algebra (1.1)-(1.2) for $\a\to0$. It contains an
$so(4)$ subalgebra of $so(5)$ generated by $\hX_i=\hJ_{4i}$ and $\hL_i=\ha\e_{ijk}\hJ_{jk}$,
while the momentum space is realized as the coset space $S^3=SO(4)/SO(3)$.

The representations of the $so(4)$ algebra can be labeled by the eigenvalues of
$\hA_i^2$, $\hL_i^2$ and $\hL_3$, where $\hA_i$ is the operator $\ha(\hL_i+\b^\mo\hX_i)$,
so that $\hX_i^2=\b^2(4\hA_i^2-\hL_i^2)$:
$$\hL_i^2\state=l(l+1)\state,\qquad\hL_3\state=m\state,$$
$$\hX_i^2\state=\b^2[4j(j+1)-l(l+1)]\,\state,\eqno(2.1)$$
with $0\le l\le2j$, $j(j+1)$ being the eigenvalue of $\hA_i^2$. The eigenvalues of $\hX_i^2$
have degeneration $2l+1$.

The momentum space can be realized as a 3-sphere, obtained by imposing the
constraint $P_i^2+P_4^2=1/\b^2$ on a four-vector. This can be shown algebraically, as in [6],
or also from a Dirac reduction of the phase space [7].
In the following, we shall mainly concentrate on the operators $\hX_i^2$ and $\hP_i^2$ and
investigate their spectra.

\subsect{2.1. The representation I}

One can define several different representations of the Snyder algebra on a Hilbert space.
Usually they are given in momentum representation.
In [2,6] it was realized by operators $\hP_i$ and $\hX_i$ defined as
$$\hP_i=P_i,\qquad\hX_i=i{\de\over \de P_i}+i\b^2P_i\(P_j{\de\over \de P_j}+\m\),\eqno(2.2)$$
$$\hL_i=-i\e_{ijk}P_j{\de\over \de P_k},\eqno(2.3)$$
which act on functions $\q(P_i)$ of the Hilbert space, with $\m$ an arbitrary real parameter
and $-\inf<P_i<\inf$.
The operators are symmetric for the measure
$${d^3P\over(1+\b^2P_i^2)^{2-\m}},\eqno(2.4)$$
provided that the functions $\q(P_i)$ go to infinity like $(1/\sqrt{P_i^2})^{\m-\ha}$.

In this \rep the operator $\hX_i^2$ reads
$$\eqalignno{\hX_i^2=&-(1+\b^2P_\r^2)^2\({\de^2\over\de P_\r^2}+{2\over P_\r}{\de\over\de P_\r}\)&\cr
&-\m\b^2\[2(1+\b^2P_\r^2)P_\r{\de\over\de P_\r}+(1+\m)\b^2 P_\r^2+3\]+{\hL_i^2\over P_\r^2}.&(2.5)}$$
For $\m=0$, the equation $\hX_i^2\f=X_i^2\f$ has eigenfunctions [6]
$$\f_{nlm}=\const.\times\sin^l\c\ C^{(l+1)}_n(\cos\c)\ Y^l_m(P_\h,P_\vf),\eqno(2.6)$$
where we have used polar \coo $P_\r$, $P_\h$, $P_\vf$ for $P_i$, and $\c=\arctan\b P_\r$.
The functions $C^{(a)}_n$ are Gegenbauer polynomials with $n$ a nonnegative integer parameter,
and $Y^l_m(P_\h,P_\vf)$ spherical harmonics.

It is easy to see that if $\m\ne0$, the eigenfunctions (2.6) are simply multiplied by $\cos^\m\c$.
The eigenvalues are of course independent of $\m$ and read
$$X_i^2=\b^2(n^2+2nl+2n+l),\eqno(2.7)$$
with $0\le l\le n$. They can easily be identified with (2.1) by setting $n=2j-l$.

The operator $\hP_i^2=\hP_\r^2$ is trivial and its spectrum extends to the real positive line.

\subsect{2.2. The representation II}

An alternative \rep is obtained [5] by defining
$$\hP_i={P_i\over\sqrt{1-\b^2P_k^2}},\qquad\hX_i=i\sqrt{1-\b^2P_k^2}\ {\de\over\de P_i},\eqno(2.8)$$
with $P_k^2<1/\b^2$.
In this \rep the operators are symmetric for the measure
$${d^3P\over\sqrt{1-\b^2P_k^2}},\eqno(2.9)$$
and the operator $\hX_i^2$ reads
$$\hX_i^2=-(1-\b^2P_\r^2){\de^2\over\de P_\r^2}-{2-3\b^2P_\r^2\over P_\r}\ {\de\over\de P_\r}
+{(1-\b^2P_\r^2)\hL_i^2\over P_\r^2}.\eqno(2.10)$$
It has eigenfunctions
$$\f_{qlm}=\const.\times\sin^l\y\,\cos\y\ P^{(1/2,l+1/2)}_{q}(\cos2\y)\ Y^l_m(P_\h,P_\vf),\eqno(2.11)$$
where $\y=\arcsin\b P_\r$, and $P^{(a,b)}_{q}$ are Jacobi polynomials with $q$ a nonnegative integer.
The eigenvalues are given by
$\b^2[(2q+l+2)^2-l(l+1)-1]$. Taking $q={n-1\over2}$, one recovers the eigenvalues (2.7).

\section{3. The nonrelativistic SdS algebra}

The representations of the operators $\hx_i$ and $\hp_i$ that satisfy the SdS algebra can be
obtained from the operators $\hX_i$ and $\hP_i$ of the Snyder algebra by taking the linear
combinations [13]
$$\hx_i=\hX_i+\l{\b\over\a}\hP_i,\qquad\hp_i=(1-\l)\hP_i-{\a\over\b}\hX_i,\eqno(3.1)$$
with inverse
$$\hP_i=\hp_i+{\a\over\b}\hx_i,\qquad\hX_i=(1-\l)\hx_i-\l{\b\over\a}\hp_i,\eqno(3.2)$$
where $\l$ is a free parameter. Representations with different values of $\l$ are related by
unitary \tran [13], therefore in the following we shall consider only the case $\l=0$.

The relation between Snyder and SdS representations can be understood by considering the
embedding of $S^3$ into $Gr(3,5)$, corresponding to the branching $SO(5)\to SO(4)$.
We recall that the vectors of $Gr(3,5)$ satisfy the constraints (1.3), while those of $S^3$ satisfy
$\b^2(P_i^2+P_4^2)=1$. Taking into account the SdS gauge constraint $\a x_5+\b p_5=0$, it is easy to
see that the combination $P_\m=p_\m+{\a\over\b}x_\m$ defined as in (3.2)  satisfies the same
constraint as the vectors of $SO(4)/SO(3)$ and then transforms as the Snyder momentum.

\subsect{3.1. The momentum representation I}

Setting $\l=0$, from (3.1) and (2.2) one obtains the \rep
$$\hx_i=i{\de\over \de P_i}+i\b^2P_i\(P_j{\de\over \de P_j}+\m\),\qquad
\hp_i=P_i-i{\a\over\b}\left[{\de\over \de P_i}+\b^2P_i\(P_j{\de\over \de P_j}+\m\)\right].\eqno(3.3)$$
As for the Snyder model, the eigenfunctions can be written in terms of
$\hx_i^2$, $\hL^2$ and $\hL_3$.

Clearly, $\hx_i^2=\hX_i^2$, and hence the equation
$$\hx_i^2\q=x_i^2\q\eqno(3.4)$$
has the same eigenfunctions (2.6) and eigenvalues (2.7) as in the Snyder model.

The calculation of the operator $\hp_i^2$ is a bit more involved. From (3.1) one has for $\l=0$,
$$\hp_i^2=\hP_i^2-{\a\over\b}\,(\hX_i\hP_i+\hP_i\hX_i)+{\a^2\over\b^2}\hX_i^2.\eqno(3.5)$$
In the  \rep of section 2.2,
$$\hX_i\hP_i+\hP_i\hX_i=2i(1+\b^2P_\r^2)P_\r{\de\over\de P_\r}+3i+i\b^2(1+2\m)P_\r^2.
\eqno(3.6)$$

From (2.5) and (3.6) follows then
$$\eqalignno{\hp_i^2=&-{\a^2\over\b^2}\[(1+\b^2P_\r^2)^2{\de^2\over\de P_\r^2}+(1+\b^2P_\r^2)
\(1+\b^2P_\r^2+i{\b\over\a}P_\r^2\){2\over P_\r}{\de\over\de P_\r}-{\hL_i^2\over P_\r^2}\]&\cr
&-{\a\over\b}\[3i+\(i\b^2-{\b\over\a}\)P_\r^2\]+\m\a^2\[2(1+\b^2P_\r^2)P_\r{\de\over\de P_\r}+
(1+\m)\b^2 P_\r^2+3-2i{\b\over\a}P_\r^2\].&\cr&&(3.7)}$$

As for the Snyder model, it is straightforward to check that the solutions with $\m\ne0$ can
simply be obtained by multiplying those with vanishing $\m$ by $\cos^\m\c$, so we consider only
the case $\m=0$.
Then, the solutions of the eigenvalue equation $\hp_i^2\f=p_i^2\f$ can be deduced from
those of (3.4) by noting that the substitution $\f=(1+\b^2P_\r^2)^{-{i\over2\a\b}}\q$ brings the
equation to the same form as (2.5), with $X_i^2\to{\b^2\over\a^2}p_i^2$, and hence its
eigenfunctions differ only by a phase from those of $\hx_i^2$:
$$\f_{nlm}=\const.\times\sin^l\c\ \cos^{i\over\a\b}\c\ C^{(l+1)}_n(\cos\c)\ Y^l_m(P_\h,P_\vf).
\eqno(3.8)$$
The operators $\hx_i^2$ and $\hp_i^2$ are therefore related by a unitary transformation,
and the eigenvalues of $\hp_i^2$ are the same as those of $\hx^2_i$, except for a multiplicative
constant:
$$p_i^2=\a^2(n^2+2nl+2n+l)\eqno(3.9)$$

This could have been predicted on the ground of the duality between $\hx_i$ and $\hp_i$.
It follows that in the SdS model also the eigenvalues of the momentum square (and hence of the
energy) are quantized and that they do not depend on $\b$.

\subsect{3.2. The momentum representation II}

Also for SdS one can adopt the alternative \rep of section 2.2 [13].
Starting from  (2.8), one obtains, for $\l=0$,
$$\hx_i=i\sqrt{1-\b^2P_k^2}\ {\de\over\de P_i},\qquad
\hp_i={P_i\over\sqrt{1-\b^2P_k^2}}-i{\a\over\b}\sqrt{1-\b^2P_k^2}\ {\de\over\de P_i}.\eqno(3.10)$$

As before, for $\l=0$, the operator $\hx_i^2$ coincides with $\hX_i^2$ and its eigenfunctions
and eigenvalues are given respectively by (2.11) and (2.7).

The operator $\hp_i^2$ reads instead
$$\hp_i^2=-{\a^2\over\b^2}\[(1-\b^2P_r^2){\de^2\over\de P_r^2}+
{2-\left(3\b^2+2i{\b\over\a}\right)P_r^2\over P_r}\ {\de\over\de P_r}\]
-{(1+2i\a\b)P_r^2+3i{\a\over\b}\over(1-\b^2P_r^2)}
+{\a^2\hL^2\over\b^2P_r^2}.\eqno(3.11)$$
This result had been obtained in [13] for a slightly different operator.

In analogy with the calculations done in the previous section, the eigenvalue equation for $\hp_i^2$
can be
reduced to the form (2.10) by introducing a function $\q$ such that $\f=(1-\b^2P_r^2)^{i/2\a\b}\q$.
The solution is therefore
$$\f_{qml}=\const.\times\sin^l\y\ \cos^{1+i/2\a\b}\y\ P_q^{(l+\ha,\ha)}(\cos2\y)\ Y^l_m(P_\h,P_\vf),
\eqno(3.12)$$
with $\y=\arcsin\b P_\r$. As in the Snyder case, taking $q={n+1\over2}$, one recovers the eigenvalues
(3.9).

\subsect{3.3. The position representations}

The duality of the SdS algebra for interchange of $\hx_i$ with $\hp_i$ permits to define
position representations by simply exchanging the roles of the phase space coordinates.
Alternatively, such representations can be obtained starting from those
of the symmetries of $S^3$ in Beltrami coordinates and using \tran analogous to (3.1).

From (3.3) and (3.10) one obtains in this way the action of
the momentum and position operators on the Hilbert space of functions of $X_i$.
We report them in the case $\l=0$, where they read, respectively,
$$\hp_i=i{\de\over \de X_i}+i\a^2X_i\(X_k{\de\over \de X_k}+\m\),\qquad
\hx_i=X_i-i{\b\over\a}\left[{\de\over \de X_i}+\a^2X_i\(X_k{\de\over \de X_k}+\m\)\right],$$
and
$$\hp_i=i\sqrt{1-\a^2X_k^2}\ {\de\over\de X_i},\qquad
\hx_i={X_i\over\sqrt{1-\a^2X_k^2}}-i{\b\over\a}\sqrt{1-\a^2X_k^2}\ {\de\over\de X_i}.$$

Position representations can be useful in some problems, like the hydrogen atom, where
the potential is a nontrivial function of $X_i$.

\section{4. Conclusions}
We have investigated some properties of the nonrelativistic SdS algebra, and in particular
its Hilbert space representations. This algebra is notable because it generates deformed
commutation relations without breaking the Lorentz invariance. Since position and momentum
are related by a duality, their operators have identical spectra, except for a
multiplicative constant, and both are discrete due to the compactness of the algebra.
Of course many other unitary equivalent representations exist besides the ones
considered in this paper, that lead to the same physical results.

Our discussion may be useful for the study of nontrivial systems in SdS space. At present,
only the free particle and the harmonic oscillator have been discussed [13]. An interesting
system to investigate would be the hydrogen atom. However, preliminary calculations seem to
indicate that it leads to third order differential equations, that are difficult to study.

As suggested in ref.\ [6] for the flat limit, our results may also be employed for the
construction of a quantum field theory on curved Snyder space, by exploiting the lattice-like
structure that has emerged from our investigation.

\beginref
\ref [1] J. Kowalski-Glikman and L. Smolin, \PR{D70}, 065020 (2004).
\ref [2] H.S. Snyder, \PR{71}, 38 (1947).
\ref [3] S. Mignemi, \PL{B672}, 186 (2009).
\ref [4] J. Kowalski-Glikman and S. Nowak, \IJMP{D12}, 299 (2003).
\ref [5] S. Mignemi, \PR{D84}, 025021 (2011).
\ref [6] Lei Lu and A. Stern, \NP{B854}, 894 (2011).
\ref [7] F. Girelli, T. Konopka, J. Kowalski-Glikman and E.R. Livine, \PR{D73}, 045009 (2006).
\ref [8] F. Girelli and E. Livine, \JHEP{1103}, 132 (2011);
\ref [9] E.J. Hellund and K. Tanaka, \PR{94}, 192 (1954);
 J.M. Romero and A. Zamora, \PR{D70}, 105006 (2004);
J.M. Romero, J.D. Vergara and J.A. Santiago, \PR{D75}, 065008 (2007);
 R. Banerjee, S. Kulkarni and S. Samanta, \JHEP{0605}, 077 (2006);
 M.V. Battisti and S. Meljanac, \PR{D79}, 067505 (2009); \PR{D82}, 024028 (2010);
C. Leiva,  Pramana {\bf 74}, 169 (2010);
C. Leiva, J. Saavedra and J.R. Villanueva, Pramana {\bf 80}, 945 (2013);
P.G. Castro, R. Kullock and F. Toppan, \JMP{52}, 062105 (2011);
Lei Lu and A. Stern, \NP{B860}, 186 (2012);
 S. Pramanik and S. Ghosh, \IJMP{A28}, 1350131 (2013);
S. Pramanik, S. Ghosh and P. Pal, \PR{D90}, 105027 (2014);
J.M. Lorenzi, R. Montemayor and L.F. Urrutia, \arx{1306.2996};
M.A. Gorji, K. Nozari and B. Vakili \PR{D89}, 084072 (2014);
 G. Amelino-Camelia and V. Astuti,  \arx{1404.4773}.
\ref [10] S. Mignemi, \AdP{522}, 924 (2010).
\ref [11] M. Born, \RMP{21}, 463 (1949).
\ref [12] S. Mignemi, \CQG{26}, 245020 (2009).
\ref [13] S. Mignemi, \CQG{29}, 215019 (2012).
\ref [14] C. Chryssomakolos and E. Okon, \IJMP{D13}, 1817 (2004);
H.G. Guo, C.G. Huang and H.T. Wu, \PL{B663}, 270 (2008);
 M.C. Carrisi and S. Mignemi, \PR{D82}, 105031 (2010);
 R. Banerjee, K. Kumar and D. Roychowdhury, \JHEP{1103}, 060 (2011);
B. Iveti\'c, S. Meljanac and S. Mignemi, \CQG{31}, 105010 (2014);
 S. Mignemi and R. \v Strajn, \PR{D90}, 044019 (2014);
 M.M. Stetsko, \JMP{56}, 012101 (2015).
\ref [15] C.N. Yang, \PR{72}, 874 (1947).
\ref [16] R. Gilmore, {\it Lie groups, Lie algebras and some of their applications}, Wiley 1974.
\ref [17] P.A.M. Dirac, {\it Lectures on quantum mechanics}, Yeoshua University, New York 1964.

\endref
\end